\documentclass[10pt,twocolumn]{revtex4}
\usepackage{graphicx}
\usepackage{amsmath}
\usepackage{amssymb}
\usepackage{color}
\usepackage{mathrsfs}

\begin{document}

\title{Molecular rotation movie filmed with high-harmonic generation}

\author{Yanqing He$^{1,\ast}$, Lixin He$^{1,}$\footnote{These authors contributed equally to this work.}}

\author{Pengfei Lan$^{1}$}\email{pengfeilan@mail.hust.edu.cn}
\author{Baoning Wang$^1$}
\author{Liang Li$^1$}
\author{Xiaosong Zhu$^1$}
\author{Wei Cao$^1$}

\author{Peixiang Lu$^{1,2,}$}\email{lupeixiang@mail.hust.edu.cn}

\affiliation{%
$^1$Wuhan National Laboratory for Optoelectronics and School of Physics,
Huazhong University of Science and Technology, Wuhan 430074,
China\\
$^2$Hubei Key Laboratory of Optical Information and  Pattern Recognition, Wuhan Institute of Technology, Wuhan 430205, China\\}

\date{\today}

\begin{abstract}

Direct imaging of molecular dynamics is a long-standing goal in physics and chemistry. As an emerging tool, high-harmonic spectroscopy (HHS) enables accessing molecular dynamics on femtosecond to attosecond time scales. However, decoding information from the harmonic signals is usually painstaking due to the coherent nature of high-harmonic generation (HHG).
Here we show that this obstacle can be effectively overcome by exploiting machine learning in HHS.
Combining the machine learning with an angle-resolved HHS method, we demonstrate that the rich dynamics of molecular rotational wave packet is fully reconstructed from the angular distributions of HHG measured at various time delays of the probe pulse.
The experimental retrievals are in good agreement with the numerical simulations. These findings provide a comprehensive
picture of molecular rotation in space and time which will facilitate the development of related researches on molecular dynamics imaging.

\end{abstract}                         %

\maketitle

\section{Introduction}

Resolving molecular dynamics in time and space, videlicet, making a ``molecular movie'', has been long desired for its potential in revealing the intermediate processes of chemical reactions and biological transformations \cite{film1,film2,film3}. This enticing prospect, however, has encountered grand challenges over the last century due to the formidable spatiotemporal scales of molecular dynamics. Coherent diffraction imaging techniques based on X-ray and electron beam have enabled identifying the atomic positions in molecules with sub-{\aa}ngstrom spatial resolution \cite{xray1,xray2,xray3,xray4,xray5}. However, the achievable temporal resolutions of these methods are currently limited to few tens to hundreds of femtoseconds.

Recent advances in strong-field physics suggest that electrons liberated from a molecule can be driven back by the oscillating electric field to revisit its parent ion \cite{three1,three2,QRS1}. Such an encounter gives rise to a number of electron-ion collisional phenomena and offers new opportunities for the probe of molecular dynamics \cite{phe1,phe2,phe3,phe4,phe5,phe6,phe7,phe8}.
In particular, high-harmonic generation (HHG) from molecules provides a snapshot of molecular structure and dynamics
at the recombination instant of the returning electrons. The information is encoded in the amplitudes, phases, and polarization of the harmonic fields. Extracting information from HHG signals systematically is known as high-harmonic spectroscopy (HHS), which recently has been an emerging tool for ultrafast molecular detection. In HHS, the temporal resolution comes from the interval between ionization
and recombination events, and the spatial resolution is related to the de Broglie wavelength of the
recombining electron, which promises resolving molecular
dynamics on attosecond time scale \cite{d1,d2,d3,d4,d5} and imaging
molecular struture with {\aa}ngstrom spatial resolution \cite{o1,o2,o3}.
However, despite these superiorities, HHS also encounters rigorous challenges in the information decoding from the HHG signals.
Especially for the complex-valued molecular information, of which both the amplitude and phase are coherently embroiled in the harmonic spectrum, the decoding is rather arduous or even unattainable due to the poor ability of the conventional iterative algorithm commonly-used in HHS in dealing with the phase problem (complex-valued data) \cite{o3,wx}.
Thus in many previous works on HHS, the phase information of the dynamics is either left out \cite{ky,vm} or addressed under certain approximations \cite{ren}, which makes it very challenging to make a full molecular movie.

In recent years, with the emergence of artificial intelligence, the modern machine learning community has developed techniques with remarkable abilities to recognize, classify, and characterize complex sets of data with a high degree of accuracy, which have been widely applied in diverse domains, e.g., image
processing \cite{ml1,ml2}, face recognition \cite{ml3}, optical tomography \cite{ml4}, and playing the game Go \cite{ml5}, etc. More recently, machine learning has been explored as a tool in genetics \cite{eml1}, condensed-matter physics \cite{eml2}, and material science \cite{eml3}. Owing to
the excellent ability in complex data-processing, the machine learning is also expected to be a powerful candidate
for the decoding in HHS studies.

\begin{figure*}[t]
\centerline{
\includegraphics[width=15cm]{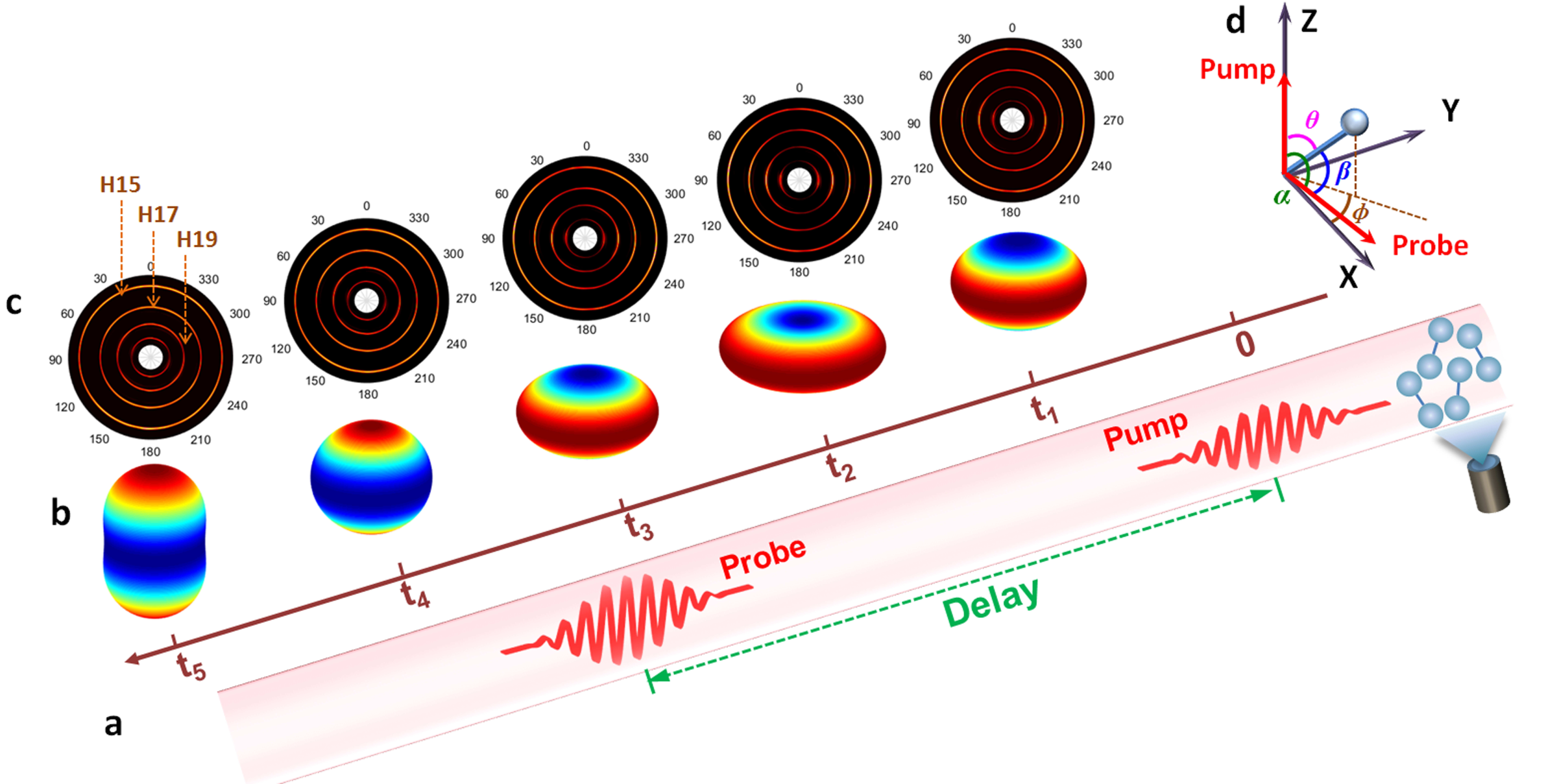}}
 \caption{\label{fig1} Sketch of the angle-resolved HHS method for molecular RWP imaging. (a) Schematic diagram of the experimental scheme. The first pump pulse polarized along z-axis is applied at t=0 to excite the molecular RWP. The subsequent probe pulse with adjustable time delay is applied to interact with the molecules to generate high-order harmonics. The harmonic ADs at each time delay are measured by scanning the polarization direction of the probe pulse in the xz-plane [as illustrated in (d)].
(b) Illustration of the 3D-RWPs simulated at different time delays. (c) Illustration of the angle-resolved harmonic spectra measured at the time delays in (b). (d) Sketch of the geometry in our experiment.
}
\end{figure*}

In the present work, we demonstrate the application of machine learning in HHS to probe molecular rotational dynamics with high resolution and fidelity. Molecular rotational control by coherently manipulating the rotational wave packet (RWP) has been proved an efficient way to connect the molecular and laboratory frames \cite{ro1,ro2,ro3,ro4,ro5}, which is significant for molecular structure imaging \cite{o1,o2,o3} and chemical reaction steering \cite{d3,chem}. In experiment, the molecular frame dynamics is usually convolved with the rotational distribution of molecules in laboratory. A throughout knowledge of molecular rotation in space and time therefore is prerequisite for resolving the molecular frame dynamics from the measurement. By using an angle-resolved HHS method, we show that the rotational dynamics of the molecular RWP excited by a pump pulse is recorded in the angular distributions (ADs) of HHG measured at various polarization directions and time delays of the probe pulse.
The time-dependent angular probability density of the RWP, is then retrieved from the measured harmonic ADs by using a machine learning algorithm, which provides a full movie of the spatiotemporal evolution of the molecular RWP. The underlying mechanism of the molecular rotational dynamics is further revealed by analyzing the populations and phases of the excited rotational states. All the experimental retrievals agree well with the numerical simulations of the time-dependent Schr\"{o}dinger equation (TDSE).

\section{Results}

\textbf{Experimental measurement of the angle-resolved HHG.}
HHG through laser-molecule interaction occurs under a much faster timescale than molecular rotation. The instantaneous rotational state of molecules can be encoded in the generated harmonic spectra. In our work, the HHG experiment is performed with a pump-probe scheme as sketched in Fig. \ref{fig1}(a).
A pump pulse with moderate intensity, linearly polarized along the z-axis [as illustrated in Fig. \ref{fig1}(d)], is first applied to the molecular ensemble to create molecular RWP through the stimulated Raman transitions \cite{ro1}. Subsequently, an intense linearly polarized probe pulse with adjustable
time delay is applied to interact with the excited molecules to generate high-order harmonics (for more experimental details, see Methods). Impulsively excited by the pump pulse, the molecules are temporarily well confined in a narrow cone around the polarization direction of the pump pulse (i.e., the z-axis). Afterwards the created RWP disperses and evolves under the field-free condition [as illustrated in Fig. \ref{fig1}(b)]. To characterize the spatial distribution of the molecular
RWP, we measure the angular distributions (ADs) of the generated harmonics by scanning the polarizing angle $\alpha$ of the probe pulse with respect to the pump pulse in
the xz plane spanned by the field vectors of these two
pulses [see the geometry in Fig. \ref{fig1}(d)].
The temporal evolution of molecular RWP is traced by
repeating this process at various time delays of the probe
pulse. The time-dependent harmonics ADs then provide a series of snapshots of the molecular RWP during the evolution [see Fig. \ref{fig1}(c)]. The measurement at each time delay will make a ``frame'' in the ``movie'' of the evolution of molecular RWP.

\begin{figure}[t]
\centerline{
\includegraphics[width=9cm]{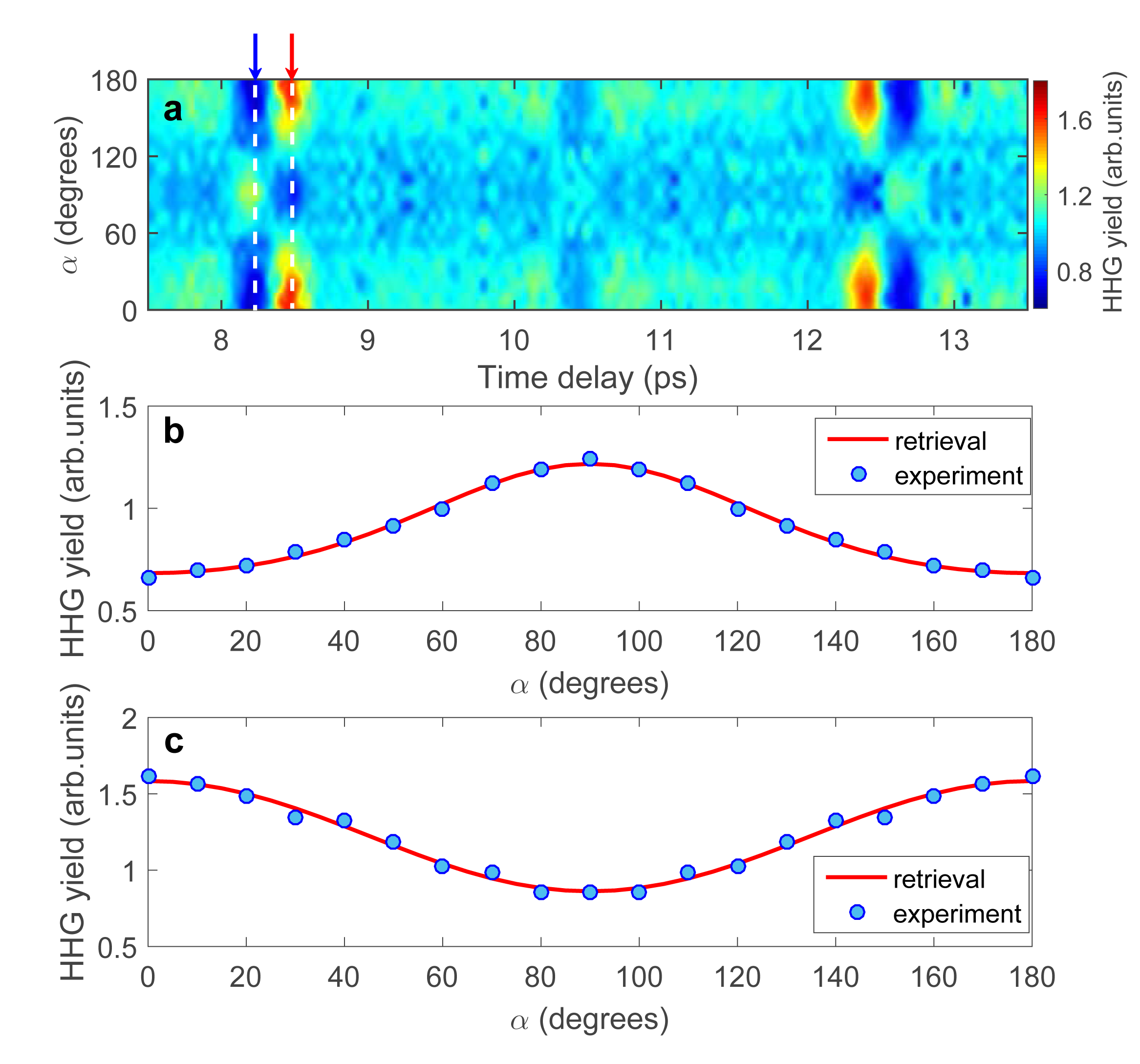}}  
 \caption{\label{fig2} Experimentally measured harmonic ADs of N$_2$. (a) The ADs of H17 as a function of the time delay of the probe pulse. Here, the harmonic yields have been normalized by the result of an unexcited ensemble (i.e., without the pump pulse). (b) and (c) are results (dots) measured at the time delay of 8.2 ps [antialignment revival, as indicated by the blue arrow in (a)] and 8.45 ps [alignment revival, as indicated by the red arrow in (a)], respectively. The red lines in (b) and (c) show the reconstructions with the retrieved MAD $\rho(\theta,t)$.
}
\end{figure}

To demonstrate our scheme, we choose the commonly-used nitrogen (N$_2$) molecule as a prototype to do the HHG experiment. Figure \ref{fig2}(a) shows the measured time-dependent ADs of harmonic 17 (H17) from N$_2$. The harmonic ADs show strong delay dependence at the rotational revivals, e.g., around t=8.4 ps (1T$_{rev}$, T$_{rev}=\frac{1}{2B_0c}$ is the revival period of N$_2$ with $B_0$ the rotational constant and $c$ the velocity of light) and 12.6 ps (1.5T$_{rev}$). In Fig. \ref{fig2}(b) and (c), we plot the results (dots) measured at t= 8.2 ps and 8.45 ps [indicated by the blue and red arrows in Fig. \ref{fig2}(a)], respectively.
One can see that the angle-dependent HHG yields measured at t=8.2 ps exhibit a maximum at $\alpha=90^\circ$. Driven by a linearly polarized probe pulse, HHG from N$_2$ is most pronounced when the laser is parallel to the
molecular axis due to the $\sigma_g$ symmetry of the highest-occupied molecular orbital (HOMO) of N$_2$ \cite{homo1,homo2}. The maximal HHG yield at $\alpha=90^\circ$ in Fig. \ref{fig2}(b) implies that most molecules are aligned perpendicularly to the pump pulse, which just corresponds to an antialignment revival at t=8.2 ps. While at t=8.45 ps, the maximal HHG yield appears at $\alpha=0^\circ$ (or $180^\circ$) where the probe pulse is parallel to the pump pulse, indicating an alignment revival at this moment. From the time-dependent harmonic ADs, the rapid quantum transformation of the molecular RWP from antialignment to alignment is directly identified. However, it's important to note that the measured harmonic AD is not an honest presentation of the distribution of molecular RWP, since the orientation-dependent electronic response is also convoluted in the HHG signals. For instance, at the same revivals (alignment/antialignment), HHG from CO$_2$ molecule shows different ADs from N$_2$ (see Supplementary Note 2). To gain a molecular movie of the evolution of molecular RWP, a further decoding from the measured harmonic signals is required.

\begin{figure}[t]
\centerline{
\includegraphics[width=9cm]{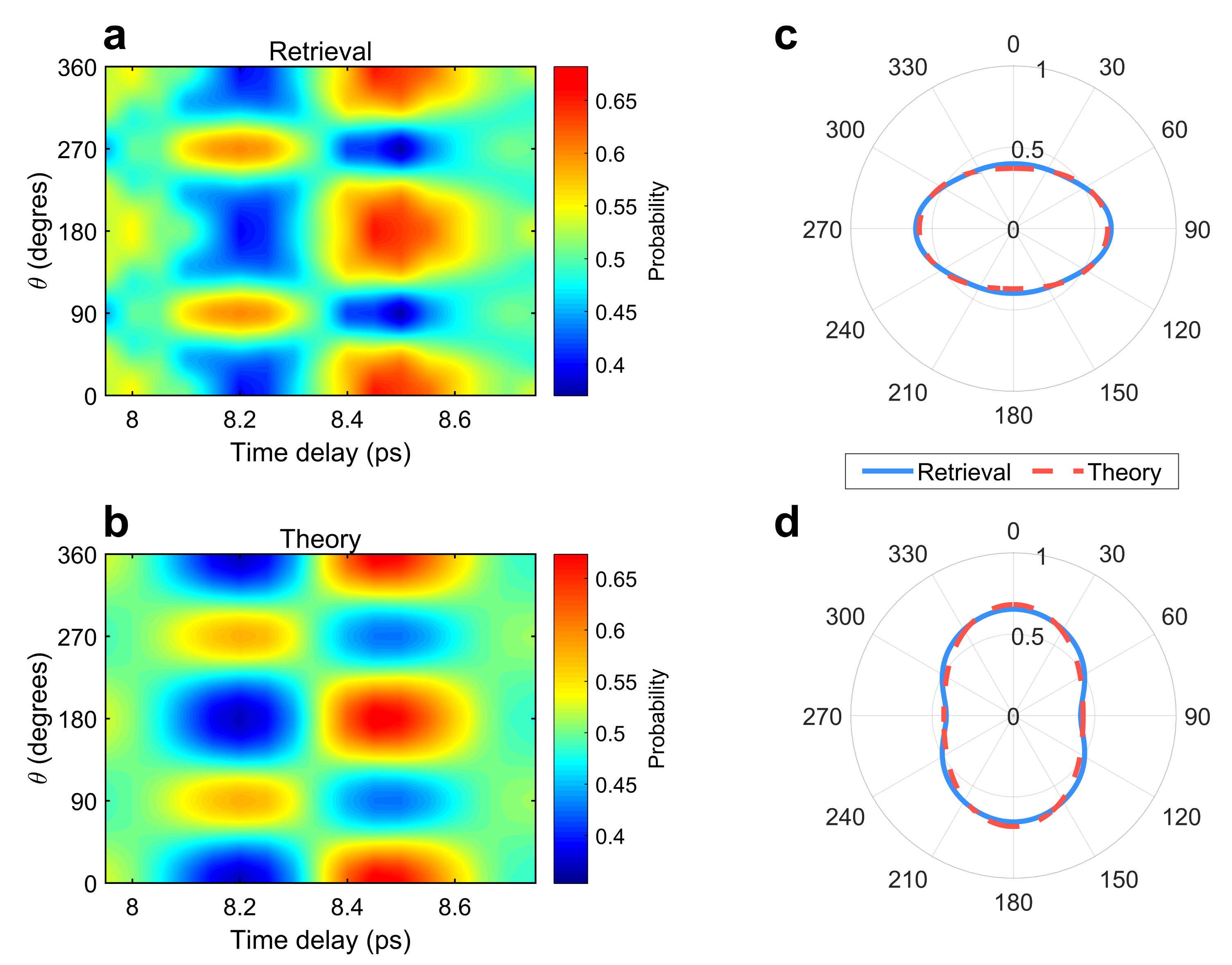}}  
 \caption{\label{fig3} Comparison of the experimentally retrieved and theoretically calculated MADs. (a) The MAD  $\rho(\theta,\phi,t)$ retrieved from the experiment. (b) same as (a), but for the calculation result. Excited by the linearly polarized pump laser, the MAD $\rho(\theta,\phi,t)$ is $\phi$-independent. (a) and (b) just for the result of $\phi$=0$^\circ$. (c) and (d) show the polar plots of the retrieved (solid line) and calculated (dashed line) MADs $\rho(\theta,\phi,t)$ at the time delay of 8.2 ps (antialignment revival) and 8.45 ps (alignment revival), respectively. }
\end{figure}

\textbf{Recuperation of molecular rotation movie from the angle-resolved HHG signals.}

After the excitation of the pump pulse, the evolution of the RWP created from each initially populated rotational state $|JM\rangle$ [the eigenstate of the field-free rigid rotor described by the spherical harmonic $Y_J^M(\theta,\phi)$] can be expanded as \cite{njp}
\begin{eqnarray}
\Psi_{JM}(\theta,\phi, t)=\sum\limits_{J'M'}C_{JM,J'M'}(t)Y_{J'}^{M'}(\theta,\phi),
\end{eqnarray}
where $C_{JM,J'M'}(t)=|C_{JM,J'M'}(t)|e^{i\varphi_{JM,J'M'}(t)}$ is the time-dependent complex coefficient of each rotational eigenstate with the amplitude $|C_{JM,J'M'}(t)|$ and phase $\varphi_{JM,J'M'}(t)$. The labels $J$ (or $J'$) and $M$ (or $M'$) are associated with the total angular momentum and its projection onto the space-fixed z axis.
Assuming a thermal distribution of the initial rotational states, the time-dependent molecular axis distribution (MAD) $\rho(\theta,\phi, t)$, which directly reflects the molecular rotational dynamics after the interaction of the molecular ensemble with the pump pulse, can be written as a weighted average of the modulus squares of the wave packet $\Psi_{JM}(\theta,\phi,t)$ \cite{ro1,njp,tdse1,tdse2,tdse3,qrs2}, i.e.,
\begin{eqnarray}
\rho(\theta,\phi,t)=\sum\limits_{JM}\Gamma_{JM}|\Psi_{JM}(\theta,\phi,t)|^2,
\end{eqnarray}
where $\Gamma_{JM}$ is the statistical weight (i.e., the population) of the initial state $|JM\rangle$ given by the Boltzmann distribution. HHG from the impulsively excited molecular ensemble can be related to the single-molecule contribution through the time-dependent MAD $\rho(\theta,\phi, t)$. Considering the coherent nature of HHG, the time-dependent harmonic ADs will thus be given by \cite{tdse3,qrs2}:
\begin{eqnarray}
I_q(\alpha,t)=\left|\int_{\phi=0}^{2\pi}\int_{\theta=0}^{\pi}
S_q(\beta)\rho(\theta,\phi,t)\sin\theta  d\theta d\phi \right|^2.
\end{eqnarray}
Here $\theta$ and $\phi$ are the polar and azimuthal angles of molecular
axis with respect to the z and x axes [see the geometry in Fig. \ref{fig1}(d)]. $\beta$ is the angle between the molecular axis and the polarization of the probe laser, which obeys $\cos\beta=\sin\theta\sin\alpha\cos\phi+\cos\theta\cos\alpha$.
$S_q(\beta)$ is the induced dipole moment of the $q$-th harmonic related to the
single molecule response with given orientation. In our reconstruction,
we assume that the single molecule dipole moment $S_q(\beta)$ is known from the calculations by the quantitative rescattering (QRS) theory \cite{QRS1,qrs2}, of which the accuracy and validity in modeling molecular HHG have been well established. Equations (1)-(3) set up a function between the harmonic ADs $I_q(\alpha,t)$ and the complex coefficients $C_{JM,J'M'}(t)$. The reconstruction of molecular rotation movie requires the retrieval of the complex coefficients (both amplitude and phase) for all the relevant initial states from the measured harmonic ADs. To this end, we introduce the machine learning to the HHS study, and employ a global optimization algorithm---the simulated annealing (SA) algorithm (see Supplementary Note 1), which is an efficient solver for approximating the global optimum of a given function in a large search space, to do the retrieval.

\begin{figure}[t]
\centerline{
\includegraphics[width=8cm]{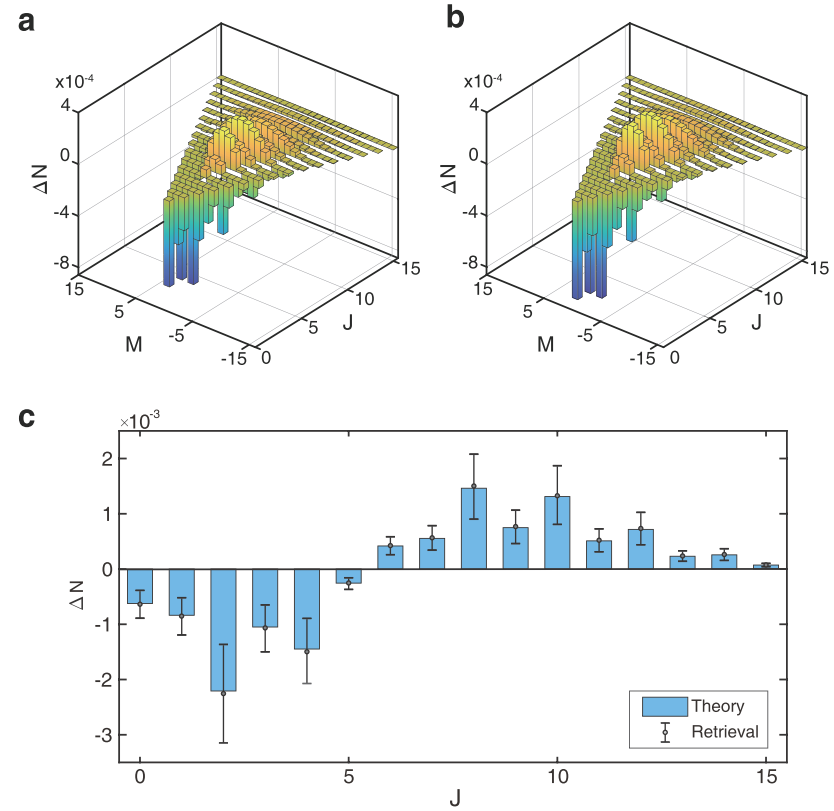}}  
 \caption{\label{fig4} Population change of the rotational states induced by the pump pulse. (a) The change of the  rotational-state population retrieved from the experiment data.
(b) Same as (a), but for the calculation results. (c) Population change of the $J$ states obtained by summing up the individual $M$ state contributions. The bars and dots are for the calculated and retrieved results, respectively. Here, the error bars represent the standard deviation of the retrievals at different time delays.}
\end{figure}

In Fig. \ref{fig2} and \ref{fig3}, we have recuperated the time-dependent MAD $\rho(\theta,\phi,t)$ as well as the harmonic ADs $I_q(\alpha,t)$ in terms of Eq. (2) and (3) by using the retrieved coefficients $C_{JM,J'M'}$. As shown in Fig. \ref{fig2} (b)-(c), the recuperated harmonic ADs (solid lines) are in excellent agreement with experiment data (dots), indicating a high accuracy of the SA algorithm. In Fig. \ref{fig3} (a), we show the recuperated MAD $\rho(\theta,\phi,t)$. Note that imposed by the linearly polarized pump pulse, the resulting RWP has a cylindrical symmetry in space, thus $\rho(\theta,\phi,t)$ is independent on the azimuthal angle $\phi$ \cite{ro1,njp}. Figure \ref{fig3} just shows the result of $\phi=0^\circ$.
From Fig. \ref{fig3} (a), one can see clear signatures of the antialignment and alignment revivals at $t$=8.2 and 8.45 ps, respectively. To evaluate the quality of the retrievals, we have theoretically calculated the MAD $\rho(\theta,\phi,t)$ by solving the TDSE of molecular RWP (see Methods). In our calculations, the laser intensity of the pump pulse has been optimized so that the squared difference between the measured and
calculated signals is minimal. The calculated results are shown in Fig. \ref{fig3} (b). It's evident that the main distribution of the experimental retrieval agrees well with the theoretical one.
For better comparison, the retrieved (solid line) and calculated (dashed line) results at the time delays of 8.2 ps and 8.45 ps are also presented in Fig. \ref{fig3} (c)-(d). The overall agreement between the retrievals (solid line) and calculations (dashed line) are excellent.

A deep insight of this rich rotational dynamics is further revealed by analyzing the amplitude and phase of the RWP created from each initial rotational states. The amplitude is associated with the populations of the rotational states. Excited by the pump pulse, the populations of the contributing rotational states are redistributed due to the transitions between different states.
The rotational transition induced by the pump pulse can be quantified by analyzing the
population change $\Delta N_{JM}=N_{JM}-\Gamma_{JM}$ of each rotational state. Here, $N_{JM}$=$\sum\limits_{J'M'}\Gamma_{J'M'}|C_{J'M',JM}|^2$ represents the population of the rotational state $|JM\rangle$ after the excitation of the pump pulse.
Figure \ref{fig4}(a) depicts the $\Delta N$ retrieved from the experiment. Note that after the impulsive excitation of the pump pulse, the created RWP evolves field-freely. The population $N_{JM}$ is almost time-delay-independent.
The result in Fig \ref{fig4}(a) is an average of the retrievals at different time delays. It's clear that excited by the pump pulse, the lower-lying rotational states are transferred to much higher rotational levels. Owing to the linear polarization of the pump pulse, the magnetic quantum number $M$ is conserved in the transitions, i.e., $\Delta M=0$ \cite{ro1,njp}. Since the lower-lying rotational states possess smaller magnetic quantum number $M$, the populations of high rotational states are mainly concentrated at small $M$ states. Moreover, the distribution
over the magnetic quantum number $M$ after the interaction is symmetric, agreeing with the cylindrical symmetry of the excitation.
These results are in consistent with the theoretical calculations shown in Fig \ref{fig4}(b). For a direct comparison
with the theory, we count up the populations of the sub-$M$ states for each $J$ state. As shown in Fig. \ref{fig4}(c), regardless of the experimental error, the agreement between the experiment (dots) and theory (bars) is excellent.

\begin{figure}[t]
\centerline{
\includegraphics[width=8.5cm]{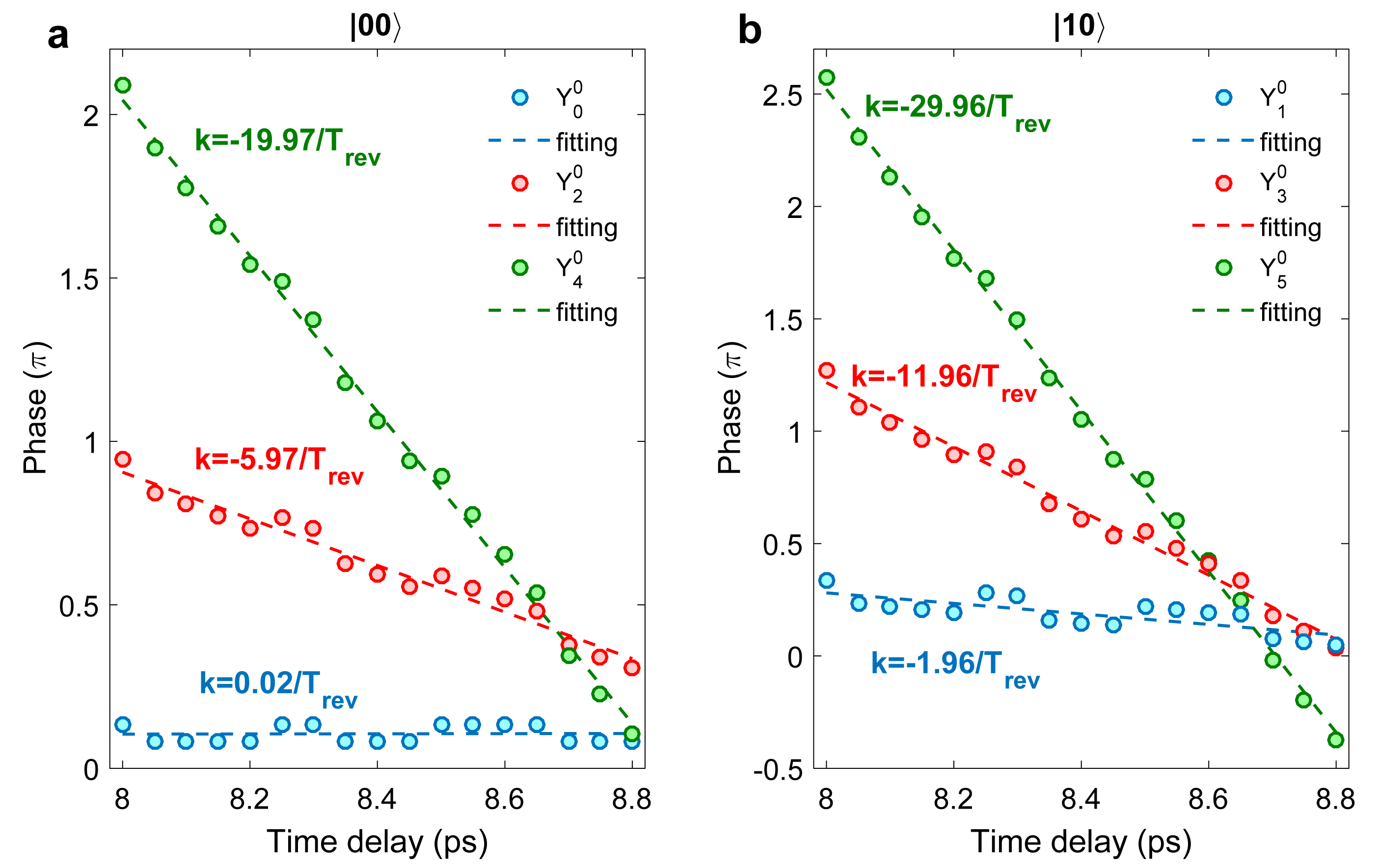}}  
 \caption{\label{fig5} Time-dependent phases of the coherent rotational states created from different initial states. (a) Time-dependent phases (dots) of the rotational eigenstates Y$_0^0$, Y$_2^0$, and Y$_4^0$ retrieved for the initial state $|00\rangle$. Dashed lines show the linear fitting of the retrievals. (b) Same as (a), but for the rotational eigenstates Y$_1^0$, Y$_3^0$, and Y$_5^0$ from the initial state $|10\rangle$.
}
\end{figure}

\begin{figure*}[t]
	\centerline{
		\includegraphics[width=17cm]{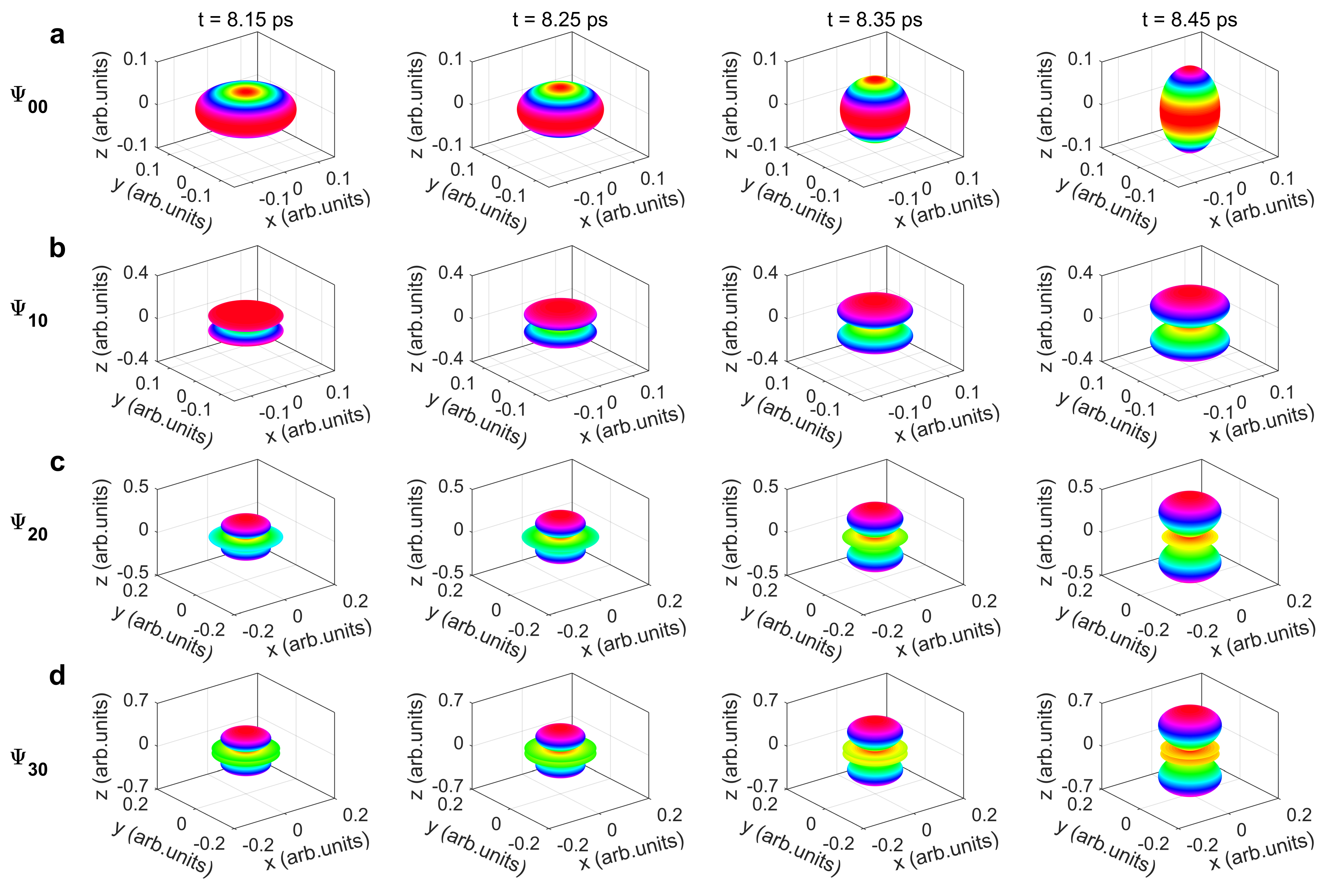}}  
	\caption{\label{fig6} Reconstructions of the RWPs created from different initial states. (a) Spatiotemporal evolution of the RWP reconstructed for the initial state $|00\rangle$. (b)-(d) Same as (a), but for the initial states $|10\rangle$, $|20\rangle$, and $|30\rangle$, respectively.
	}
\end{figure*}

In Fig. \ref{fig5}, we show the retrieved phase $\varphi_{JM,J'M'}$ of the relevant rotational states as a function of the pump-probe delay.
Figure \ref{fig5}(a) depicts the results of the rotational eigenstates Y$_0^0$, Y$_2^0$, and Y$_4^0$ created from the initial $|00\rangle$ state.
It's obvious that the retrieved phase of each rotational eigenstate (dots) exhibits a nearly linear change with the time delay.
Dotted curves in Fig. \ref{fig5} show the linear fitting of these retrievals.
The slopes (as indicated) of the fitting curves are very close to $-\frac{E_J}{\hbar}$, where $E_J=2\pi\hbar B_0cJ(J+1)=\pi\hbar J(J+1)/T_{rev}$ is the eigen energies of the considered rotational eigenstates. This result is consistent with the field-free nature of the evolution of the RWP after the interaction, which is described by a phase factor $e^{-i\frac{E_J}{\hbar}t}$ \cite{ro1,njp}.
With the complex coefficients $C_{JM,J'M'}$ retrieved, we can then reproduce the dynamics of the RWP created from each initial rotational state in terms of Eq. (1). Figure \ref{fig6} (a)-(d) display some selected snapshots of the RWP created from the $|00\rangle$ , $|10\rangle$, $|20\rangle$, and $|30\rangle$ states, respectively. One can see that, owing to the time-dependent phase beating of the coherently populated rotational eigenstates, the RWPs created from different initial rotational states at the rotation revivals are synchronously condensed/stretched along the z axis, thus leading to the oblate/prolat angular distribution of molecules at the antialignment/alignment revival. Through a thermal average over these created RWPs, a complete movie of molecular rotational dynamics then is achieved (see the movie in the Supplemental Material).

\section{Discussion}

In summary, we demonstrate a full experimental reconstruction of molecular rotational dynamics by using an angle-resolved HHS method in combination with the machine learning algorithm. Under the excitation of a pump pulse, a coherent molecular RWP is created in experiment, which afterwards is probed by
measuring the time-dependent ADs of HHG through the interaction the molecular ensemble with a subsequent probe pulse.
By utilising a machine learning algorithm---the SA algorithm, the time-dependent angular probability distribution of the molecular RWP is successfully retrieved from the measured harmonic ADs. A full movie of the evolution of molecular RWP in space and time is then achieved. The rich dynamics of molecular rotation is further illuminated by analyzing the populations and phases of the
excited rotational states. All the experiment results are consistent with the TDSE simulations.
The present results afford a thoroughgoing
knowledge of the spatiotemporal evolution of the field-free molecular RWP, which is extremely useful for future researches on ultrafast imaging of strong-field molecular frame dynamics.

On the other hand, machine learning provides a practical and robust method for extracting information from the HHG spectra. It's general and can be effectively extended to other molecules, like CO$_2$ (see Supplementary Note 2) and even some more complex polyatomic molecules. Moreover, HHG from molecules also contains abundant information on the electronic structure and dynamics. The exploitation of machine learning in HHS could also open access to the ultrafast charge migration in molecules. In the future, the machine learning is expected to be an extremely effective technique in attosecond physics for actually ``watching" a chemical reaction.

\section{Methods}

\textbf{Experimental methods.} Our experiment is carried out by using a commercial Ti: sapphire laser system (Astrella-USP-1K, Coherent, Inc.), which delivers 35 fs,
800 nm pulses at a repetition rate of 1 kHz. The output laser is spilt by
a beam splitter to produce the pump and probe pulses. The pump
pulse polarized along the z-axis is first used to create molecular RWP. Subsequently, the intense probe pulse with adjustable
time delay is applied to interact with the molecule ensemble to generate high-order harmonics.
A motorized delay lines and a half-wave plate are installed in the arm
of the probe pulse to adjust its time delay and polarization with respect to the pump one.
These two beams are focused to a gas jet (500 $\mu$m diameter) by a 300-mm focal-length
lens. The background gas pressure is maintained at 20 Torr and
the rotational temperature is estimated to be about 100 K \cite{temp}. The generated harmonic
spectrum is detected by a homemade flat-field soft x-ray spectrometer \cite{he1,he2}.

\textbf{Theoretical methods.} To compare with the experiment, we have also simulated the molecular rotational dynamics by solving the time-dependent Schr\"{o}dinger equation (TDSE) of the molecular RWP. For each initial rotational state $|JM\rangle$, the TDSE reads \cite{tdse1,tdse2,tdse3,qrs2}
\begin{eqnarray}
i\frac{\partial\Psi_{JM}(\theta,\phi,t)}{\partial t}& = &[B_0\textbf{J}^2-\frac{E^2(t)}{2}(\alpha_{\parallel} cos^2\theta    \nonumber\\
& & +\alpha_{\perp} sin^2\theta)]\Psi_{JM}(\theta,\phi,t).
\end{eqnarray}
Here, \textbf{J} is the rotation operator, $B_0$ is the rotational constant of N$_2$ molecule, $\alpha_{\parallel}$ and $\alpha_{\perp}$ are the  the anisotropic polarizabilities in parallel and perpendicular directions
with respect to the molecular axis, respectively. $E(t)$ is the envelope of the
field vector of the pump pulse. Eq. (4) is solved with the split-operator method for each initial rotational state $|JM\rangle$ \cite{split}. Assuming a Boltzmann distribution of the rotational levels at the initial time, the MAD at each time delay can be obtained in terms of Eq. (2). The harmonic spectrum is then calculated according to Eq. (3) in the main text.
In our simulations, the single molecule response is calculated based on the QRS theory \cite{QRS1,qrs2}. The intensity of the probe pulse is estimated from the harmonic cutoff in the experiment, which is about $2.3\times10^{14}\ \mathrm{W/cm}^2$. The intensity of the pump pulse is determined by minimizing the squared difference
between the measured and calculated HHG signals, which is about $1.1\times10^{13}\ \mathrm{W/cm}^2$. Other parameters are the same as the experimental conditions.

\section{Acknowledgements}
The authors gratefully acknowledge Xu Wang for helpful
discussions. This work was supported by National Natural Science
Foundation (NSFC) of China (Grants No. 11627809,
No. 11874165, No. 11704137, and No. 11774109), Fundamental
Research Funds for the Central Universities
(2017KFXKJC002), Program for HUST Academic Frontier
Youth Team.

\section{Author contributions}
P.X.L. and P.F.L. conceived and designed the experiment. L.X.H., Y.Q.H., B.N.W., and P.F.L. performed the experiments. Y.Q.H., L.X.H., and P.F.L. performed the simulations. L.L., X.S.Z., and W.C. participated in the discussions.



\begin{thebibliography}{99}
	
	
    \bibitem{film1} Ihee, H., Lobastov, V. A., Gomez, U. M., Goodson, B. M., Srinivasan, M., Ruan, C. Y. \& Zewail, A. H. Direct imaging of transient molecular structures with ultrafast diffraction. {\it Science} \textbf{291}, 458 (2001).

	\bibitem{film2} Zewail, A. H. \& Thomas, J. M. 4D Electron Microscopy: Imaging in Space and Time (Imperial College Press, 2009).

	\bibitem{film3} Smirnova, O. \& Ivanov, M. Towards a one-femtosecond film. {\it Nat. Phys.} \textbf{6}, 159 (2010).


	\bibitem{xray1} Seibert, M. M. et al. Single mimivirus particles intercepted and imaged with an
X-ray laser. {\it Nature} \textbf{470}, 78-81 (2011)

	\bibitem{xray2} Chapman, H. et al. Femtosecond X-ray protein nanocrystallography. {\it Nature} \textbf{470},
73-77 (2011).

	\bibitem{xray3} Ihee, H. et al. Direct imaging of transient molecular structures with ultrafast
diffraction. {\it Science} \textbf{291}, 458-462 (2011).

	\bibitem{xray4} Sciaini, G. \& Miller, R. J. D. Femtosecond electron diffraction: heralding the era
of atomically resolved dynamics. {\it Rep. Prog. Phys.} \textbf{74}, 096101 (2011).

	\bibitem{xray5} K\"{u}pper, J. et al. X-ray diffraction from isolated and strongly aligned gas-phase
molecules with a free-electron laser. {\it Phys. Rev. Lett.} \textbf{112}, 083002 (2014).

    \bibitem{three1} Corkum, P. B. Plasma perspective on strong ?eld multiphoton ionization. {\it Phys. Rev. Lett.} \textbf{71}, 1994-1997 (1993).

    \bibitem{three2} Lewenstein, M., Balcou, Ph., Ivanov, M., L'Huillier, A. \& Corkum, P. B. Theory of high-harmonic generation by low-frequency laser fields. {\it Phys. Rev. A} \textbf{49}, 2117 (1994).

    \bibitem{QRS1} Lin, C. D., Le, A. T., Chen, Z., Morishita, T. \& Lucchese, R. Strong-field rescattering
physics---self-imaging of a molecule by its own electrons. {\it J. Phys. B} \textbf{43}, 122001 (2010).

    %
	\bibitem{phe1} Meckel, M. et al. Laser-induced electron tunneling and diffraction. {\it Science} \textbf{320}, 1478-1482 (2008).

	\bibitem{phe2} Blaga, C. I. et al. Imaging ultrafast molecular dynamics with laser-induced electron diffraction. {\it Nature} \textbf{483}, 194-197 (2012).

	\bibitem{phe3} Xu, J. et al. Diffraction using laser-driven broadband electron wave packets. {\it Nat. Commun.} \textbf{5}, 4635 (2014).

	\bibitem{phe4} Wolter, B. et al. Ultrafast electron diffraction imaging of bond breaking in di-ionized acetylene. {\it Science} \textbf{354}, 308-312 (2016).

	\bibitem{phe5} Huismans, Y. et al. Time-resolved holography with photoelectrons. {\it Science} \textbf{331}, 61 (2011).

	\bibitem{phe6} Meckel, M. et al. Signatures of the continuum electron phase in molecular strong-field photoelectron holography. {\it Nat. Phys.} \textbf{10}, 594 (2014).

	\bibitem{phe7} He, M. et al. Direct Visualization of Valence Electron Motion Using Strong Field Photoelectron Holography. {\it Phys. Rev. Lett.} \textbf{120}, 133204 (2018).

	\bibitem{phe8} Tan, J. et al. Determination of the Ionization Time Using Attosecond Photoelectron Interferometry. {\it Phys. Rev. Lett.} \textbf{121}, 253203 (2018)



	\bibitem{d1} Baker, S. et al. Probing proton dynamics in molecules on an attosecond time scale. {\it Science} \textbf{312}, 424 (2006).

	\bibitem{d2} Kraus, P. M. et al. Measurement and laser control of attosecond charge migration in ionized iodoacetylene. {\it Science} \textbf{350}, 790 (2015).

	\bibitem{d3} W\"{o}rner, H. J. et al. Following a chemical reaction using high-harmonic interferometry. {\it Nature} \textbf{466}, 604 (2010).

	\bibitem{d4} Li, W. et al. Time-Resolved Dynamics in N$_2$O$_4$ Probed Using High Harmonic Generation. {\it Science} \textbf{322}, 1207-1211 (2008).

	\bibitem{d5} Lan, P. et al. Attosecond Probing of Nuclear Dynamics with Trajectory-Resolved High-Harmonic Spectroscopy. {\it Phys. Rev. Lett.} \textbf{119}, 033201 (2017).

	\bibitem{o1} Itatani, J. et al. Tomographic imaging of molecular orbitals. {\it Nature} \textbf{432}, 867 (2004).

	\bibitem{o2} Haessler, S. et al. Attosecond imaging of molecular electronic wavepackets. {\it Nat. Phys.} \textbf{6}, 200 (2010).

	\bibitem{o3} Vozzi, C. et al. Generalized molecular orbital tomography. {\it Nat. Phys.} \textbf{7}, 822 (2011).



	\bibitem{wx} Wang, X. et al. Theory of retrieving orientation-resolved molecular information using
time-domain rotational coherence spectroscopy. {\it Phys. Rev. A} \textbf{96}, 023424 (2017).

	\bibitem{ky}  Yoshii, K., Miyaji, G. \& Miyazaki, K. Retrieving angular distributions of high-order harmonic generation from a single molecule. Phys. Rev. Lett. 106, 013904 (2011).

	\bibitem{vm} Makhija, V. et al. Orientation Resolution through Rotational Coherence Spectroscopy. {\it arXiv:}1611.06476 (2016)

	\bibitem{ren} Ren, X. et al. Measuring the angle-dependent photoionization cross section of nitrogen using high-harmonic generation. {\it Phys. Rev. A} \textbf{88}, 043421 (2013).


	\bibitem{ml1} Sinha, A., Lee, J., Li, S. \& Barbastathis, G. Lensless computational imaging through deep learning. {\it Optica} \textbf{4}, 1117-1125 (2017).

	\bibitem{ml2} Goy, A., Arthur, K., Li, S. \& Barbastathis, G. Low photon count phase retrieval using deep learning. {\it Phys. Rev. Lett.} \textbf{121}, 243902 (2018)

    \bibitem{ml3} Taigman, Y., Yang, M., Ranzato, M. \& Wolf, L. Closing the gap to human-level performance in face verification. deepface. {\it In IEEE Computer Vision and Pattern Recognition (CVPR)} (2014).

	\bibitem{ml4} Kamilov, U. S., Papadopoulos, I. N., Shoreh, M. H., Goy, A., Vonesch, C., Unser, M. \& Psaltis, D. Learning approach to optical tomography. {\it Optica} \textbf{2}, 517 (2015).

	\bibitem{ml5} Silver, D. et al. Mastering the game of Go with deep neural networks and tree search. {\it Nature} \textbf{529}, 484-489 (2016).




	\bibitem{eml1} Libbrecht, M. W. \& Noble, W. S. Machine learning applications in genetics and genomics. {\it Nat. Rev. Genet.} \textbf{16}, 321-332 (2015).

	\bibitem{eml2} Carrasquilla, J. \& Melko, R. G. Machine learning phases of matter. {\it Nat. Phys.} \textbf{13}, 431-434 (2017).

	\bibitem{eml3} Raccuglia, P. et al. Machine-learning-assisted materials discovery using failed experiments. {\it Nature} \textbf{533}, 73-76 (2016).


\bibitem{ro1}  Stapelfeldt, H. \& Seideman, T. Colloquium: Aligning molecules with strong laser pulses. {\it Rev. Mod. Phys.} \textbf{75}, 543 (2003).
\bibitem{ro2} Ghafur, O. et al. Impulsive orientation and alignment of quantum-state-selected NO molecules. {\it Nat. Phys.} \textbf{5}, 289-293 (2009).
\bibitem{ro3} Ohshima, Y. \& Hasegawa, H. Coherent rotational excitation by intense nonresonant laser fields. {\it Int. Rev. Phys. Chem.} \textbf{29}, 619-663 (2010).

\bibitem{ro4} Lin, K. et al. All-optical field-free three-dimensional orientation of asymmetric-top molecules. {\it Nat. Commun.} \textbf{9}, 5134 (2018).

\bibitem{ro5} Karamatskos, E. T. et al. Molecular movie of ultrafast coherent rotational dynamics. {\it arXiv:}1807.01034 (2018).


\bibitem{chem} Zare, R. N. Laser control of chemical reactions. {\it Science} \textbf{279}, 1875 (1998).


\bibitem{homo1} McFarland, B. K. et al. High harmonic generation from multiple orbitals in N$_2$. {\it Science} \textbf{322}, 1232 (2008).

\bibitem{homo2} Le, A. T., Lucchese, R. R. \& Lin, C. D. Uncovering multiple orbitals influence in high harmonic generation from aligned N$_2$. {\it J. Phys. B} \textbf{42}, 211001 (2009).


\bibitem{njp}Fleischer, S. et al. Controlling the sense of molecular rotation. {\it New J. Phys.} \textbf{11}, 105039 (2009).

\bibitem{tdse1} Lin, K. et al. Visualizing molecular unidirectional rotation. {\it Phys. Rev. A} \textbf{92}, 013410 (2015).

\bibitem{tdse2} Lin, K. et al. Echoes in Space and Time. {\it Phys. Rev. X} \textbf{6}, 041056 (2016).

\bibitem{tdse3} He, L. et al. Real-Time Observation of Molecular Spinning with Angular High-Harmonic Spectroscopy. {\it Phys. Rev. Lett.} \textbf{121}, 163201 (2018).

\bibitem{qrs2} Le, A. T. et al. Quantitative rescattering theory for high-order harmonic generation from molecules. {\it Phys. Rev. A} \textbf{80}, 013401 (2009).


\bibitem{temp} Yoshii, K., Miyaji, G., \& Miyazaki, K. Measurement of molecular rotational temperature in a supersonic gas jet with high-order harmonic generation. {\it Opt. Lett.} \textbf{34}, 1651 (2009).

\bibitem{he1} He, L. et al. Monitoring ultrafast vibrational dynamics of isotopic molecules with frequency modulation of high-order harmonics. {\it Nat. Commun.} \textbf{9}, 1108 (2018).

\bibitem{he2} He, L. et al. Spectrally resolved spatiotemporal features of quantum paths in high-order-harmonic generation. {\it Phys. Rev. A} \textbf{92}, 043403 (2015).

\bibitem{split} Saugout, S., Charron, E. \& Cornaggia, C. H$_2$ double ionization with few-cycle laser pulses. {\it Phys. Rev. A} \textbf{77}, 023404 (2008).

\end{thebibliography}
\end{document}